
\input phyzzx
\unnumberedchapters

%
%
\def\a{\alpha}
\def\b{\beta}
\def\d{\delta}
\def\ga{\gamma}
\def\m{\mu}
\def\n{\nu}
\def\r{\rho}
\def\s{\sigma}
\def\ee{\epsilon}
\def\gm{\Gamma}
\def\La{\Lambda}
%
%
\def\WW{\scriptscriptstyle R}
\def\BB{\scriptscriptstyle B}
\def\VV{\scriptscriptstyle V}
\def\AA{\scriptscriptstyle A}
\def\HH{\scriptscriptstyle H}
\def\SCS{S_{\rm CS}}
\def\cv{c_{\VV}}
\def\RR{{\rm I\!\!\, R}}
\def\idx{\int \! d^3\!x\>}

%

\mathsurround=2pt
\rightline{FTUAM 93/03}
\rightline{LPTHE 93/06}
\rightline{NIKHEF-H 93-03}
\date{}


\REF\Jones{V.F.R. Jones, Ann. Math. {\bf126} (1987) 335.}
\REF\Homfly{P. Freyd, D. Yetter, J. Hoste, W.B.R. Lickorish, K. Millet
and A. Ocneau, Bull. Am. Math. Soc. {\bf 12} (1985) 239.}
\REF\Witten{E. Witten, Commun. Math. Phys. {\bf 121} (1989) 351.}
\REF\Guada{E. Guadagnini, M. Martellini and M. Mintchev, Nucl.
Phys. {\bf 330B} (1990) 557.
\nextline
P. Cotta-Ramusino, E. Guadagnini, M. Martellini and M. Mintchev,
Nucl. {\bf 330B} (1990) 577.}
\REF\Dunne{For a canonical analysis in which quantization is prior
to reduction see G.V. Dunne, R. Jackiw and C.A. Trugenberger, Ann.
Phys. {\bf 194} (1989) 197.}
\REF\Physical{J.D. Lykken, J. Sonnenschein and N. Weiss, Int. J. Mod.
Phys. A 6 29 (1991) 5155.\nextline
T.T. Burwick. A.H. Chamseddine and K.A. Meissner, Phys. Lett.
{\bf B284} (1992) 11.}
\REF\Guadareg{E. Guadagnini, M. Martellini and M. Mintchev, Phys.
Lett. {\bf 227} (1989) 111.}
\REF\Gmr{G. Giavarini, C.P. Martin and F. Ruiz Ruiz, Nucl. Phys.
{\bf 381B} (1992) 222.}
\REF\BRSinvariant{L. Alvarez-Gaum\'e, J.M.F. Labastida and A.V.
Ramallo, Nucl. Phys. {\bf 334B} (1990) 103. \nextline
M. Asorey and F. Falceto, Phys. Lett. {\bf 241} (1990) 31.}
\REF\Freed{ D.S. Freed and R.E. Gompf, Phys. Rev. Lett. {\bf 66}
(1991) 1255, Comm. Math. Phys. {\bf 141} (1991) 79.}
\REF\Epstein{H. Epstein and V. Glaser, Ann. Inst. Henri Poincar\'e
{\bf XIX} (1973) 211.}
\REF\Collins{J.C. Collins, {\it Renormalization} (Cambridge
University Press, Cambridge, 1984).}
\REF\Jaffe{A. Jaffe and A. Lesniewski, in {\it Non-perturbative
quantum field theory}, edited by G. 't$\!$Hooft, A. Jaffe, G. Mack,
P.K. Mitter and R. Stora (Plenum Press, New York, 1988).}
\REF\Blasi{A. Blasi and R. Collina, Nucl. Phys. {\bf B345} (1990)
472.\nextline
F. Delduc, C. Lucchesi, O. Piguet and S.P. Sorella, Nucl. Phys.
{\bf B346} (1990) 313.}
%


\titlepage

\title{{\seventeenbf The universality of the shift of the Chern-Simons
parameter for a general class of BRS invariant regularizations}}

{\author{ G. Giavarini$^\star$}}\foot{Address after May 1, 1993: {\it
INFN Gruppo collegato di Parma and Dipartimento di Fisica
dell'Uni\-ver\-sit\`a di Parma, Viale delle Scienze, I-43100 Parma,
Italy.}}
\vskip-.8truecm
\address{{\it Laboratoire de Physique Th\'eorique et Hautes Energies,
              Universit\'es Paris VII \nextline
              Tour 14-24, $5^{eme}$ \'etage,
              2 Place Jussieu, 75251 Paris Cedex 05, France}}

\author{ C. P. Martin}
\address{{\it Departamento de F\'\i sica Te\'orica,  C-XI,
              Universidad Aut\'onoma de Madrid,\nextline
              Cantoblanco, Madrid 28049, Spain}}

\author{ F. Ruiz Ruiz}
\address{{\it NIKHEF-H, Postbus 41882, 1009 DB Amsterdam,
              The Netherlands}}

\vskip 1 true cm

\noindent
We consider a biparametric family of BRS invariant regularization
methods of $SU(N)$ Chern-Simons theory (the parameters defining the
family taking arbitrary values in $\RR^2$) and show that the shift
$k\to k +\,{\rm sign}(k)\, N$ of the Chern-Simons parameter $k$
occurs for arbitrary values of the family defining parameters.
This supports irrefutably the conjecture that the shift of
$k$ is universal for BRS invariant regulators.

\endpage

\pagenumber=2

The interest of perturbative quantization of $SU(N)$ Chern-Simons
theory lies on the possibility of finding a series expansion
of the Jones Polynomial [\Jones] and of its generalizations
[\Homfly], as well as of other topological invariants of
three-manifolds, \eg\ the partition function [\Witten]. This
program would procure a definition of topological invariants by
means of techniques widely used in quantum field theory. It is
worth mentioning along this line the discovery [\Guada] of integral
representations for the coefficients of the HOMFLY polynomial
[\Homfly]. There are other reasons that support perturbative
quantization of Chern-Simons theory; among them, the following
two. First, perturbative quantization
provides an alternative to canonical quantization as used in Ref.
[\Witten], where reduction to the physical degrees of freedom is
prior to quantization [\Dunne]. It is important to emphasize in
this regard that there is no a priori reason why perturbative
quantization and canonical quantization as used in Ref. [\Witten]
should yield the same quantum theory so that
agreement is not guaranteed, least trivial. Secondly, the
perturbative framework constitutes a very useful tool for studying
models with Chern-Simons gauge fields coupled to matter [\Physical].

An important issue currently subject to debate in perturbative
quantization of the theory is the meaning of radiative corrections
to the Chern-Simons parameter and their dependence on the
renormalization scheme one chooses (Ref. [\Guadareg] versus
Refs. [\Witten,\Gmr,\BRSinvariant]).
The purpose of this paper is to provide incontestable evidence
that one-loop radiative corrections to the classical Chern-Simons
parameter $k$ are universal for all gauge invariant regularization
methods, thus validating the position defended in Refs.
[\Witten,\Gmr,\BRSinvariant]. The latter radiative corrections give
for $k$ the shift [\Gmr,\BRSinvariant]
$$
k\to k+\,{\rm sign}(k) \, \cv  \, ,
\eqn\shift
$$
where $\cv$ denotes the
quadratic Casimir operator in the adjoint representation of the
gauge group [$\cv\!=\!N$ for $SU(N)$]. Here we do not
attempt to give an explanation (still to be found) of this
highly non-trivial universality but to exhibit it. We would like
to stress that if the parameter $k$ has the same meaning
in perturbative quantization as in canonical quantization, the
shift in eq. \shift\ is a necessary condition for the semiclassical
evaluation of the partition function [\Freed] to agree with the
non-perturbative value as computed with surgery techniques
[\Witten].

To better understand what we mean, let us recall very briefly some
well known results from standard local perturbative renormalization
[\Epstein,\Collins].
Let us consider a quantum field theory that is renormalizable
by power counting and let us assume that there exists a {\it
regularization method} that explicitly preserves the symmetries of
the theory. If we denote by $\phi_{\BB}$ and $k_{\BB}$
the bare fields and the bare parameters of the regularized quantum
theory, perturbative renormalization ensures that there always exist
dimensionless functions $Z_\phi$ and $Z_k$ of the regulator such that
the Green functions of the fields $\phi_{\WW} =Z^{-1}_\phi \phi_{\BB}$
as functions of the parameters $k_{\WW} = Z^{-1}_k k_{\BB}$ remain
finite and satisfy the Ward identities as the regulator is removed.
The fields $\phi_{\WW}$ and the parametres $k_{\WW}$ are called
renormalized fields and renormalized parameters and are finite as the
regulator is removed; the functional that generates finite renormalized
Green functions being called the renormalized effective action $\gm$.
It is well known that the requirement of finite renormalized Green
functions does not in
general fix $Z_\phi$ nor $Z_k$, for any further symmetry-preserving
finite renormalization still gives rise to finite renormalized Green
functions. To eliminate this ambiguity one has to choose what is called
a {\it renormalization scheme}, while preserving the symmetries of the
theory. This is as far as one can go with standard local perturbative
renormalization. Our conjecture following Ref. [\Jaffe] is precisely
that for Chern-Simons theory there is a natural choice for $Z_k$,
namely
$$
Z_k=1 \,.
\eqn\scheme
$$
This choice corresponds to taking $k_{\WW}=k_{\BB}\equiv k$, which
is the parameter entering in eq. \shift. Prior to the discussion
of the reasons why we say the choice $Z_k=1$ is natural is the
observation that $Z_k=1$ is only valid for finite theories and not for
merely renormalizable theories (such as QCD), since for the latter
ones $Z_k=1$ does not take care of UV divergences. We recall in
this regard that Chern-Simons theory is finite [\Blasi].

Of course, for the choice $Z_k=1$ to make sense, all regularization
methods preserving gauge invariance should give the same corrections
to the parameter $k$. If this is the case, the parametrization
eq. \scheme\ is natural in the sense that gauge invariance, if
preserved at the regularized level, fixes all ambiguities introudced
by regularization. We would like to observe at this point that eq.
\shift\ holds for all gauge invariant regularization methods used
as yet [\Gmr,\BRSinvariant] if $Z_k=1$, and remind that the latter
shift \shift\ is a necessary condition for the partition function
evaluated non-perturbatively using surgery techniques [\Witten] to
agree with its semiclassical expression [\Freed]. The parametrization
$Z_k=1$ is also natural in the sense that the value of $k_{\WW}$ in
terms of $k_{\BB}$ needs not to be calculated at each order in
perturbation theory. In the sequel we show that
one-loop radiative corrections to the parameter $k$ are the same
for a biparametric family of gauge invariant regularization methods,
pattern which has been observed for isolated gauge invariant regulators
[\Witten,\Gmr,\BRSinvariant]. It is important to realize that
uniqueness of quantum corrections to $k$ does not mean that
the renormalized effective action is unique. As a matter of fact,
there are infinitely many different renormalized effective actions,
each of them corresponding to a different choice of renormalization
constants $Z_\phi$ for the fields. What happens is that wave-function
renormalizations preserving BRS invariance are cohomologically trivial
in the BRS sense and hence do not contribute to gauge invariant
radiative corrections. Let us now move on to the description of our
calculation.

The Chern-Simons classical action is given by
$$
\SCS = -\,{ik\over 4\pi} \idx \ee^{\m\n\r}
    \left( {1\over 2}\,\, A^a_\m \partial_\n A^a_\r
  + {1 \over 3!}\,f^{abc}A^a_\m A^b_\n A^c_\r  \right) \>,
$$
with obvious notation and $k$ for convenience chosen to be positive.
Our regularization method consists in adding to the classical action
a higher covariant derivative term so that the action becomes
$$
\eqalign{
S_\La = \SCS &
+ {k\over 4\pi}\idx \left[\, {u\over 4\La} \> F^a_{\m\n}F^{a\,\m\n}
      - {iv \over 2\La^2} \> \ee^{\m\n\r}
                   F^a_{\m\s} (D_\n F_\r^{~\s})^a
      + {1\over 4 \La^3}\> (D_\r F_{\m\n})^a (D^\r F^{\m\n})^a
          \,\right] \cr
& + \idx \left[\,- b^a \partial A^a
                 + (J^{a\m} - \partial^\m \bar{c}^a)\, (D_\m c)^a
                 - {1\over 2}\>f^{abc} H^a c^b c^c \,\right] \>, \cr}
$$
where we have already included the standard gauge fixing term in the
Landau gauge $\partial A^a =0$. Here $u$ and $v$ are arbitrary real
parameters with $uv \ne 1$ if $u$ is negative.
As usual, the higher covariant derivative terms here added do not
completely regularize the theory, for there is a finite number
of 1PI Feynman diagrams still divergent by power counting. We
regularize the latter diagrams by using 't$\!$Hooft-Veltman's
dimensional regularization method for theories involving parity
violating objects [\Collins]. There are other dimensional
regularization prescriptions but they are either algebraically
inconsistent or violate BRS invariance (we refer the interested
reader to Ref. [\Gmr] for details). In this way we define
renormalized Chern-Simons theory in the scheme $Z_k=1$ as the
limit $\La \to \infty$ of the limit $D \to 3$ of the
dimensionally regularized theory whose classical action is $S_\La$.
The calculation of the one-loop renormalized Chern-Simons
effective action in the scheme eq. \scheme\ is then performed as
follows:
\smallskip

\noindent{\it Step 1}. We begin recalling the expression of the
local part of the renormalized effective action up to one loop
for Chern-Simons theory:
$$
\gm^{\rm local} = - {i(k+\a)\over 4\pi} \SCS
    + \idx \left\{ -b^a\partial A^a
        + \Delta \left[\> \b\,(J^{a\m} - \partial^\m \bar c^a )A^a_\m
                   - (1+\ga)\, H^a c^a \, \right] \right\} \>,
$$
$\a,\>\b$ and $\ga$ being arbitrary coefficients of order
$\hbar$ and $\Delta$ the Slavnov-Taylor operator for
the theory (see Ref. [\Gmr] for details).
There are two types of radiative corrections:
gauge invariant radiative orrections, labelled by $\a$, and
gauge dependent radiative corrections, labelled by $\b$ and $\ga$.
We are going to show that gauge invariant corrections do not depend
neither on $u$ nor $v$. However, gauge non-invariant corrections will
depend on $u$ and $v$ but, being cohomologically trivial with respect
to the Slavnov-Taylor operator, can be absorbed by a wave-function
renormalization thus not contributing to the observables. Using
that the renormalized effective action generates renormalized
1PI Green functions, the coefficients $\a$, $\b$ and $\ga$ can be
uniquely determined by computing the vacuum polarization tensor
$\Pi^{ab}_{\m\n}$, the ghost self-energy $\Omega^{ab}$ and the
$Hcc$-vertex $V^{abc}$ at one loop. To calculate the latter Green
functions we evaluate the limit $\La\to\infty$ (step 3) of the
limit $D\to 3$ (step 2) of their regularized counteparts
$\Pi^{ab}_{\m\n}(D,\La)$, $\Omega^{ab}(D,\La)$ and $V^{abc}(D,\La)$.
\smallskip

\noindent {\it Step 2}. In computing the limit $D \to 3$ we have to
properly take care of evanescent operators [\Gmr]. It turns out that
evanescent objects do not contribute to the three-dimensional
limit so the latter limit can be calculated using the
$D\!$-dimensional gauge field propagator
$$
D^{ab}_{\m\n} = {4\pi \over k} \>
                {\d^{ab} \over q^2\, P(q^2,\La;u,v)} ~
  \left[ \La^4\,(\La^2 + v q^2)\,\ee_{\m\r\n}q^\r
       + \La^3\,(u\La^2 + q^2)\, (q^2 g_{\m\n} - q_\m q_\n)
  \right] \>,
$$
where $\ee_{\m\r\n}$ in $D$ dimensions is understood in the
't$\!$Hooft-Veltman sense [\Gmr] and
$$
P(q^2,\La;u,v) = \La^2\,{(\La^2 + v q^2)}^2
               + q^2\,{(u\La^2 + q^2)}^2 \>.
$$
Note that $u$ and $v$ are such that $P(q^2,\La;u,v)$ is positive
definite for any $q^\m$ thus ensuring that power counting and
positivity of the real part of the action $S_\La$ hold. The
calculation of the limit $D\to 3$ of the Green functions we
are interested in (and of any other Green function of the
regularized theory) only involves elementary properties of
dimensional regularization [\Collins] and is performed
straightforwardly, the result being finite and $\La\!$-dependent.
This implies perturbative finiteness for the family of theories
defined by $S_\La$ and constitutes a necessary
condition for the computation of the large $\La$ limit to make
sense.
\smallskip

\noindent{\it Step 3}. We next calculate the limit $\La\to\infty$
of the three-dimensional Green functions $\Pi^{ab}_{\m\n}(\La)$,
$\Omega^{ab}(\La)$ and $V^{abc}(\La)$ obtained in step 2. The
large $\La$ limit of the Feynman integrals entering these Green
functions can be evaluated using the $m\!$-theorem in Ref. [\Gmr],
the result having the form
$$
\Pi^{ab}_{\m\n} (p) = \cv \, J(u,v)\>\ee_{\m\r\n} p^\r\,\d^{ab}
\qquad~~
\Omega^{ab}(p) = - {\cv\over k}\> I(u,v)\>p^2 \,\d^{ab}
\qquad~~
V^{abc}(p_1,p_2) =0 \>,
$$
where $p^\m,\>p_1^\m$ and $p_2^\m$ are external momenta,
$$
J(u,v) = - {2\over 3\pi} \int_{-\infty}^{\infty} \! dt \>
       { F(t^2;u,v) \over P^2(t^2,1;u,v)}
\qquad\qquad
I(u,v) = {2\over 3\pi} \int_{-\infty}^{\infty} \! dt \>
       { {u + t^2} \over P(t^2,1;u,v)}
$$
and
$$
\eqalign{
F(t^2;u,v) & = v\,t^{10} - 2\,(uv + 2\,v^3 + 4)\,t^8
             - (3\,u^2 v - 3\,uv^3 + 25\,u + 25\, v^2)\,t^6 \cr
           & - 2\,(11\, u^2 + 4\,u v^2 + 13\, v)\,t^4
             - (5\, u^3 + 13\, uv + 5)\,t^2 - 2\,u  \> .\cr}
$$
Note that $J(u,v)$ and $I(u,v)$ are non-trivial functions of $u$
and $v$. To convince oneself that they are not constants, it is
enough to take \eg\ $uv=1$ with $u$ positive to obtain
$J(1/v,v)= (7+3v^{3/2})/3(1+v^{3/2})$ and $I(1/v,v)= 2/3(1+v^{3/2})$.
Taking now into account the arbitrariness in the choice of $Z_\phi$
and introducing the notation $Z_\phi \equiv 1 + z_\phi$, it is
straightforward to derive the following values for $\a$, $\b$
and $\ga$:
$$
\a = \cv \, \left[ \>J(u,v) - 2\,I(u,v)\>\right] \qquad\quad
\b + z_{\AA} = {4\over 3}\>{\cv \over k}\,I(u,v) \qquad\quad
\ga + 2\,z_c + z_{\HH} = 0\>.
$$
We remind that $Z_\phi$, hence $z_\phi$ are finite and can take
arbitrary values as far as they preserve BRS invariance, that is
as far as they satisfy $Z_A Z_{\bar c} = Z_c Z_H$, or equivalently
$z_{\AA} + z_{\bar c} = z_c + z_{\HH}$.
To derive a more compact expression for $\a$, we evaluate $J(u,v)$
and $I(u,v)$ in terms of the roots of the polynomial $P(t^2,1;u,v)$.
We do the latter by writing $u$ and $v$ in terms of the roots,
by using residue techniques and by recalling that $P(t^2,1;u,v)$ is
positive definite so the integrand in $J(u,v)$ and $I(u,v)$ does not
have poles on the real axis. After some lengthy calculations we
finally obtain that
$$
\a = \cv  \,.
$$
In summary, the renormalized effective action in the scheme $Z_k=1$
depends in general on the parameters $u$ and $v$ but its gauge
invariant part never does. This exhibits the universality of the
one-loop shift $k \to k + \cv$ for the family of gauge invariant
regulators we have considered.

\bigskip
\noindent{\bf Acknowledgements:} GG was supported by The Commission
of the European Communities through contract SC 900376, and FRR by
the Dutch Stichting voor Fundamenteel Onderzoek der Materie. The
authors also acknowledge partial support from Comisi\'on de
Investigaci\'on Cient\'\i fica y T\'ecnica, Spain.

\refout

\end